\documentclass[aps,prl,twocolumn,groupedaddress,amsmath,amssymb]{revtex4-1}
\usepackage{ae}
\usepackage[pdftex]{graphicx}
\usepackage{amsmath}
\usepackage{amssymb}
\usepackage{epsfig, float}
\usepackage{hyperref}

\begin{document}
\title{Anomalous Contagion and Renormalization in Dynamical Networks \\
with Nodal Mobility}
\author{Pedro D. Manrique$^{1}$, Hong Qi$^{1}$, Minzhang Zheng$^{1}$, Chen Xu$^{2}$, Pak Ming Hui$^{3}$ \\ \& Neil F. Johnson$^{1}$}
\affiliation{$^1$Physics Department, University of Miami, Coral Gables, Florida FL 33126, U.S.A.\\
$^2$College of Physics, Optoelectronics and Energy, Soochow University, Suzhou 215006, China\\
$^3$Department of Physics, Chinese University of Hong Kong, Shatin, Hong Kong}
\date{\today}

\begin{abstract}
The common real-world feature of individuals migrating through a network -- either in real space or online -- significantly complicates understanding of network processes. Here we show that even though a network may appear {\em static} on average, underlying nodal mobility can dramatically distort outbreak profiles. Highly nonlinear dynamical regimes emerge in which increasing mobility either amplifies or suppresses outbreak severity. Predicted profiles mimic recent outbreaks of real-space contagion (social unrest) and online contagion (pro-ISIS support). We show that this nodal mobility can be renormalized in a precise way for a particular class of dynamical networks.

\end{abstract}
\maketitle

Significant attention among physicists has turned to problems where the dynamics of a meme or virus, plus the network on which it is spreading, co-evolve on comparable timescales -- whether online or in real space. The rich variety of important works \cite{vesp,barrat,stanley1,onnela,watts,stanley2,baron,karsai,lazer,kaski,gonc,dodds,ker,korn,murase,sornette1,prerumor,havlin,perra,Palla,us,estrada,Gleeson,Tivnan,1,2,3,4,5,6,7,8,9,10,11,11more,Karsai2} reflects the many possible choices for how the network can evolve dynamically, and hence be modeled. For example, Watts et al. consider an evolving hierarchical network \cite{watts} while Karsai et al. consider the weakness of strong ties and Zhao et al. allow entire clusters to fragment \cite{us}.

Here we turn our attention to a more common dynamical feature of everyday human behavior whereby people join, leave and can rejoin clusters of other individuals $C_1,C_2,\dots C_M$, e.g. by sporadically checking online posts for a particular social media community or re-visiting a particular cafe. The dynamical complication comes from the fact that a returning node (i.e. individual) may find a network cluster in which membership has changed very little or a lot, depending on the mobility of all other nodes (individuals). Depending on who they then meet and when, the resulting evolution of any infection at the population level may be very different. 

Our model for this co-evolution is purposely very simple (see Fig. 1) so that we can write down, and solve numerically, coupled differential equations that mirror the outcome of numerical simulations, as well as enabling some analytical analysis. Yet it turns out surprisingly to generate highly anomalous infection profiles which mimic those from recent periods of civil unrest fueled by social media, and online political activity (Figs. 2 and 3). While we do not pretend that our model provides a unique explanation of these real-world phenomena, it serves the purpose of providing a more unified view of such social network activity. Specifically, our analysis shows that even when a network appears {\em static} on average, underlying nodal mobility generates highly nonlinear behavior in the outbreak's severity (i.e. peak infection value $H$), time-to-peak (i.e. time $T_m$ from beginning of outbreak to its peak), duration $T$, and area $A$ under the profile $I(t)$. We also provide a novel renormalization scheme that can significantly reduce the complexity of this class of dynamical network problem. 

\begin{figure}
\centering
\includegraphics[width=0.8\linewidth]{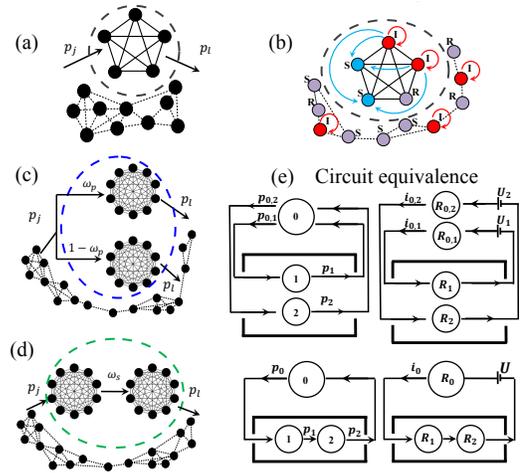}
\caption{(Color online) Our model of nodal mobility: (a) For a single, internally fully-connected cluster $C_1$ embedded in a network. Remaining links are sparse and/or weak (indicated schematically by dashed lines). Probability that a node from outside (or inside) $C_1$ joins (or leaves) the cluster on a given timestep is $p_j$ (or $p_l$). 
(b) For $C_1$ in presence of SIR (Susceptible-Infected-Recovered) process. See text for details. (c) and (d) show two-cluster case ($M=2$) in parallel and series geometries. (e) Equivalent circuits for $M=2$ clusters in parallel (top) and series (bottom).}
\label{fig1}
\end{figure}

We assume here for simplicity that the $M$ network clusters are internally fully connected (Fig. 1(a)) and that for $M>1$ the  different clusters are interconnected in simple ways, e.g. parallel (Fig. 1(c)) and series (Fig. 1(d)). Consider a network of $N$ total nodes containing a single cluster $C_1$ (Fig. 1(a)). 
At any given timestep, a node from anywhere outside $C_1$ has a
probability $p_{j}$ to join $C_1$, while a node inside $C_1$ has a probability $p_{l}$ to leave $C_1$. The number of nodes $N_{1}(t)$ in $C_1$ follows  $\dot N_{1}=-p_{l}N_{1}+p_{j}(N-N_{1})$. We focus on the case where the mean cluster size is constant so that the network appears {\em static} on average. This mean size is $\left< N_{1}\right>=Np_{j}(p_{j}+p_{l})^{-1}\equiv N\gamma_{s}$
and the sum of the mean number of nodes joining and leaving is $\mu=(N-\left<N_{1}\right>) p_j + \left<N_{1}\right> p_l \equiv N\gamma_{m}$. Hence $\gamma_{s}=p_{j}(p_{j}+p_{l})^{-1}$ characterizes the mean size of $C_1$ and $\gamma_m={2p_{l}p_{j}}(p_{l}+p_{j})^{-1}$ characterizes the {\em nodal mobility} through $C_1$. 
At any timestep, an infected agent within $C_1$ transmits a meme or virus to any susceptible within $C_1$ with probability $q_{i}$ (Fig. 1(b)). Since $C_1$ is the only fully connected cluster, we will assume that transmission from infected nodes outside $C_1$ is negligible by comparison. Since recovery is individual based, infected nodes inside and outside $C_1$ have probability
$q_{r}$ to become immune (for SIR) or susceptible again (for SIS). The infection rate $\lambda=q_{i}/q_{r}$ is the usual ratio of the infection probability to the recovery probability.

\begin{figure}
\includegraphics[width=0.8\linewidth]{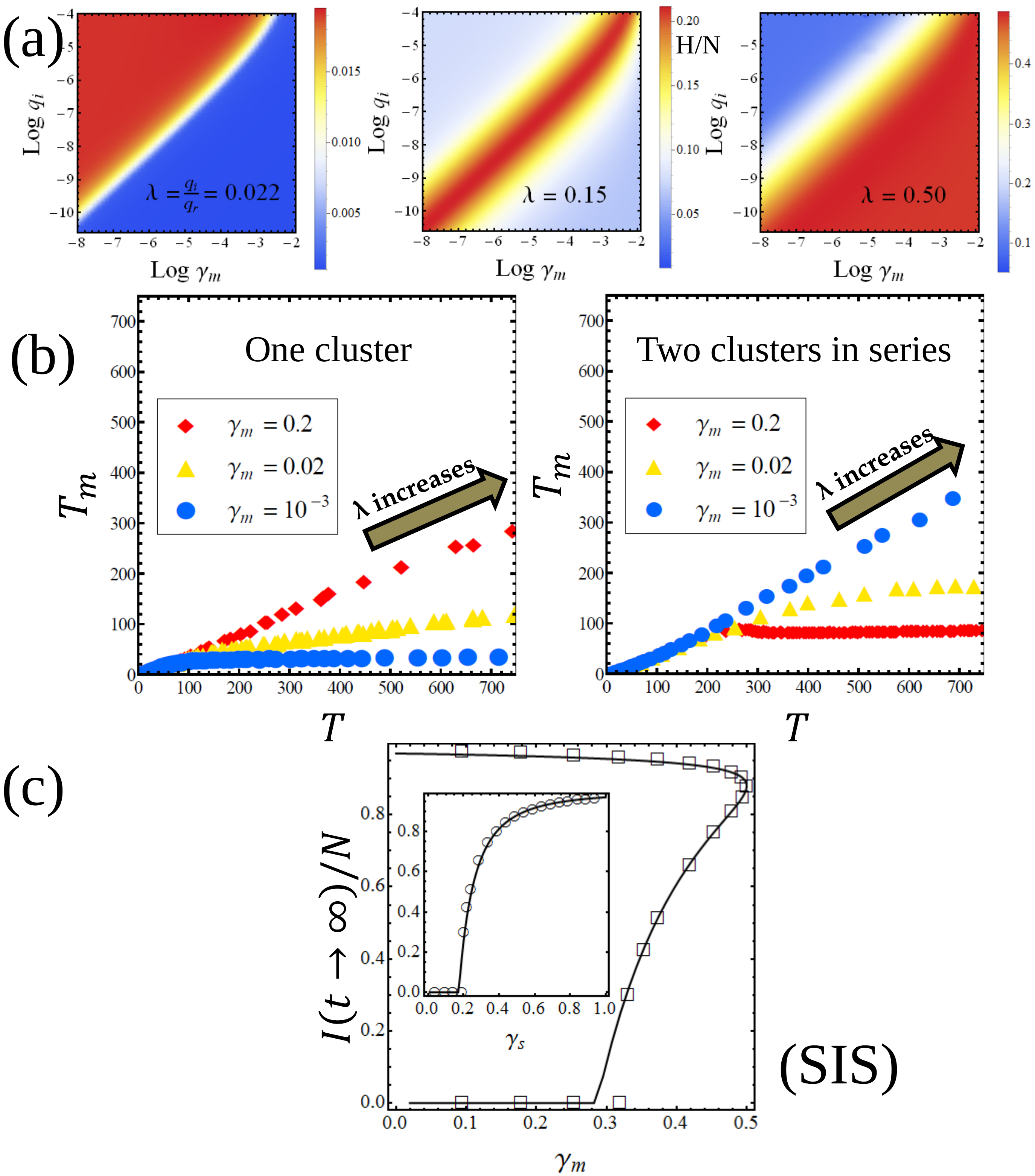}
\caption{(Color online) (a) Nonlinearity of SIR outbreak severity ($I(t)$ peak height $H$ divided by the constant $N$ which is total number of network nodes) as a function of nodal mobility $\gamma_m$ and $q_i$, for different values of the ratio $\lambda=q_i/q_r$ for one-cluster version (Fig. 1(a)). (b) Nonlinear relationship between outbreak time-to-peak $T_m$ and duration $T$ as $\lambda$ is varied. Left: one-cluster version. Right: two-cluster series version (see text). Values are averages over simulation runs. (c) SIS (Susceptible-Infected-Susceptible) for one-cluster version. Vertical scale is $I(\infty)/N$, the normalized fraction of infected nodes in the long-time limit, as a function of $\gamma_m$. $N = 1000$, $q_i = 0.0005$, $q_r = 0.015$. $p_j + p_l = 1$ for simplicity. Inset shows $I(\infty)/N$ as function of $\gamma_s$. Lines are from integrating the coupled differential equations (see SM), symbols are simulation results.}
\label{fig2}
\end{figure}

\begin{figure}
\includegraphics[width=0.8\linewidth]{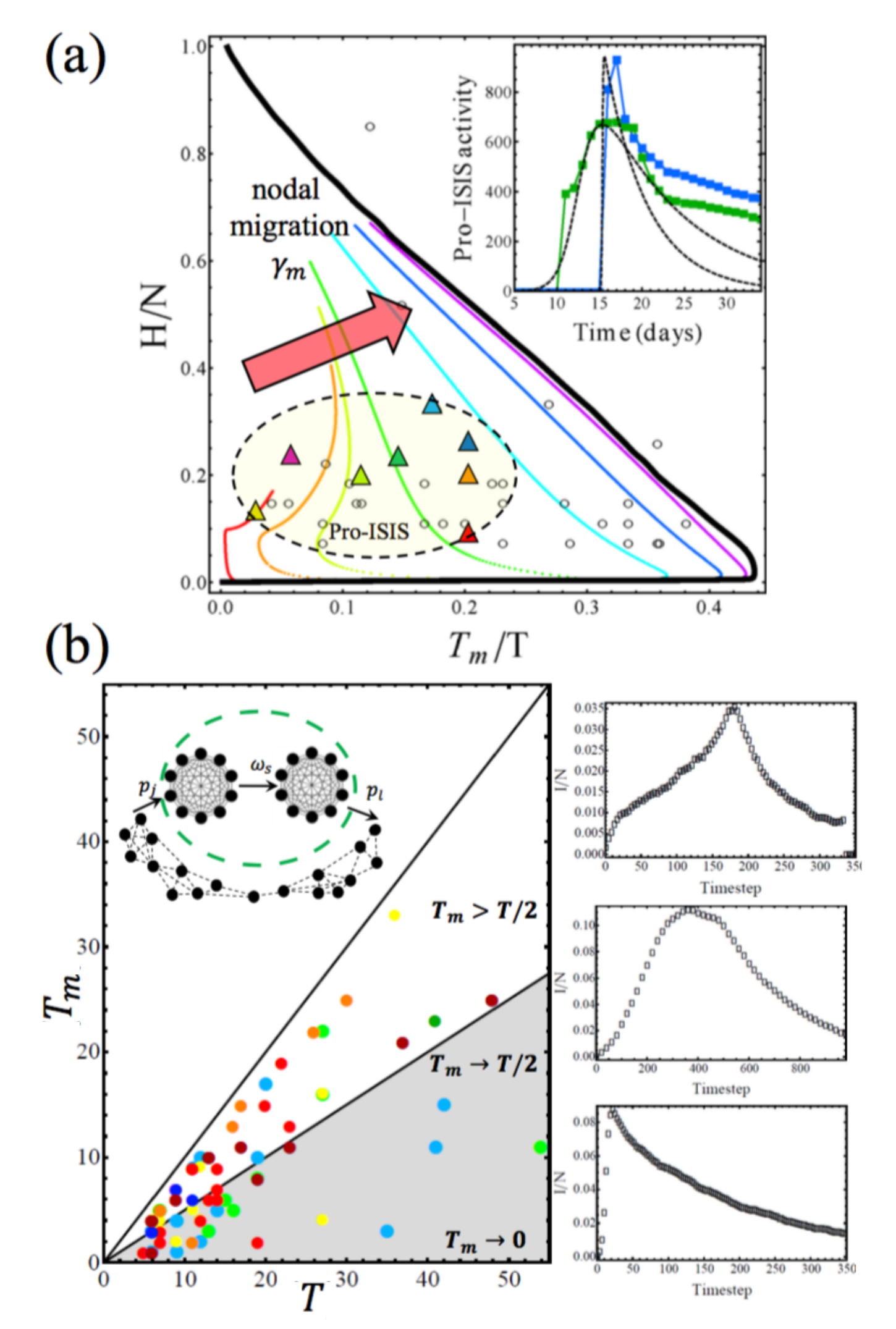}
\caption{\small{(Color online) (a) Outbreak profile descriptors $H/N$ and $T_m/T$ for one-cluster version (thin lines: Fig. 1(a)) compared to empirical data of  
on-street civil unrest (circles) \cite{IARPA} and online pro-ISIS outbreaks (colored triangles) for $T_m<T/2$. Theoretical lines obtained by integrating the coupled differential equations (see SM) for different values of nodal mobility $\gamma_m$. Thick black line shows result for standard (i.e. well-mixed) SIR model. $N=1000$, $q_i = 0.002$ throughout. Each trajectory starts near origin for $\lambda\equiv q_i/q_r=10^{-3}$ and grows until $\lambda=1$ in steps of $\delta\lambda = 10^{-3}$. Inset: Two examples of empirical profiles for online outbreaks (pro-ISIS clusters {\bf club81567093} (blue) and 
{\bf interes.publics} (green)) compared to best-fit standard SIR model. (b) Time-to-peak ($T_m$) and duration ($T$) average values for empirical civil unrest outbreaks (colored dots, see SM). Solid lines show $T_m=T$ and $T_m=T/2$ as guide. Unlike standard SIR model, two-cluster versions (e.g. inset) include range $T_m>T/2$ where many datapoints lie. Right: middle and bottom, example infection profile for one-cluster version; upper, two-cluster version.}}
\label{CSC-Ext}
\end{figure}

\begin{figure}
\includegraphics[width=0.8\linewidth]{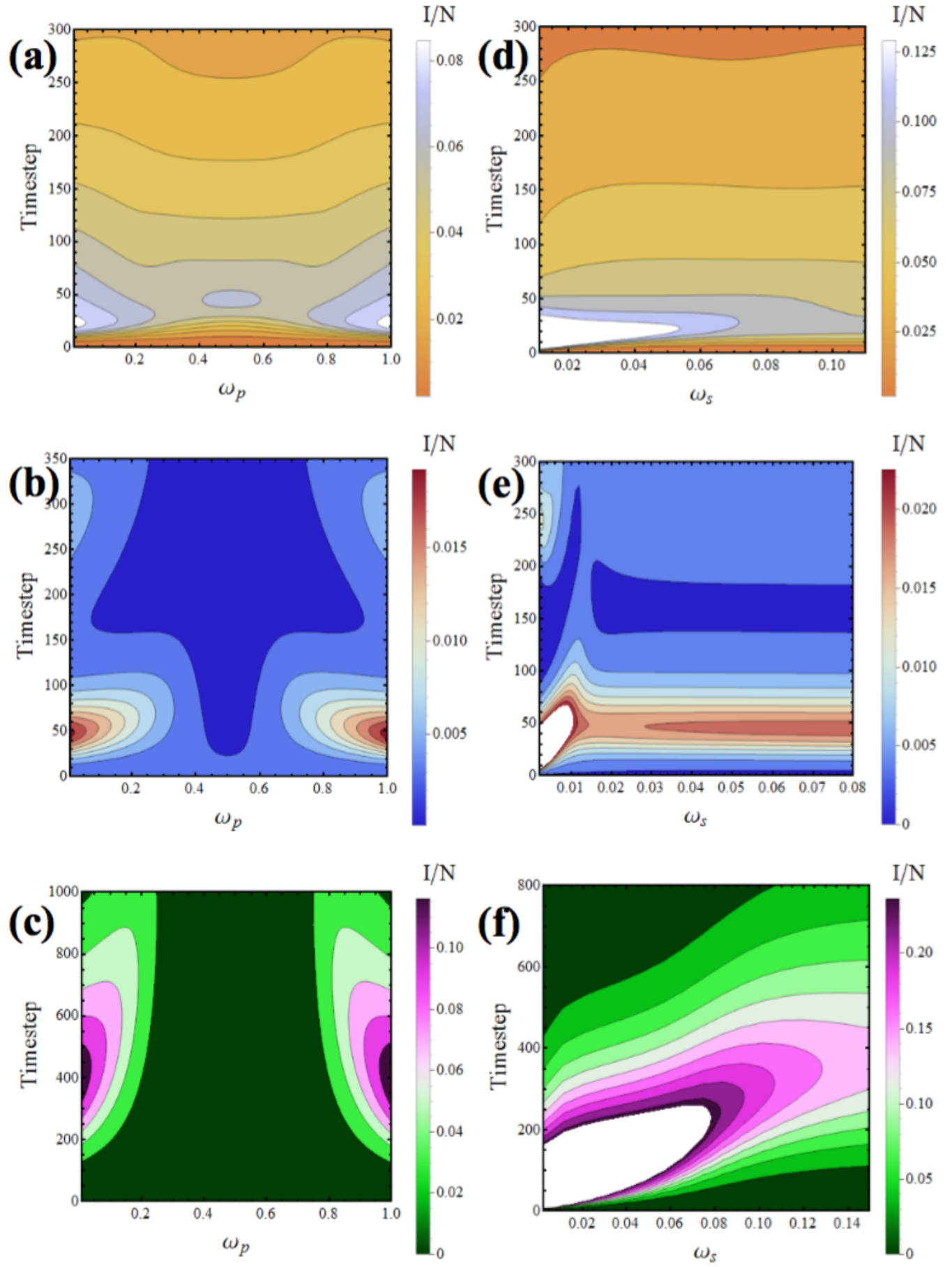}
\caption{\small{Infection profile $I(t)$ vs. time (vertical axis) for different $M=2$ cluster geometries and model probabilities (horizontal axis). (a)-(c): clusters in parallel (Fig. 1(c)).  (d)-(f): clusters in series (Fig. 1(d)). Profiles calculated by numerical integration of differential equations (see SM) for three parameter sets: (a) and (d) $q_{i}=0.005$, $\lambda=0.1$, $\gamma_{m}=0.009$; (b) and (e) $q_{i}=0.001$, $\lambda=0.1$, $\gamma_{m}=0.018$; (c) and (f) $q_{i}=0.002$, $\lambda=0.022$, $\gamma_{m}=0.0018$).}}
\label{profilesparallel}
\end{figure}

Figures 2 and 3 show that even for the one-cluster version (Fig. 1(a)), highly asymmetric and varied infection profiles emerge, with abnormally slow and/or fast decays compared to standard infection models. While it is known that models with heterogeneity in connectivity or nodal type can produce anomalous infection characteristics as compared to the usual well-mixed SIR model, our model shows this can arise in a network that appears {\em static} on average and in which the time-averaged properties of each node are the same, i.e. {\em anomalous infection profiles arise even though each node spends the same average time in cluster $C_1$ and has the same average number of links over time}. Figure 2(a) shows that for small $\lambda<0.15$, there is a monotonic nonlinear decrease of the outbreak severity with increasing nodal mobility $\gamma_m$. This might be expected since spending less time in the cluster exposes an individual (i.e. mobile node) to less risk of infection. However, one could imagine a competing mechanism whereby increased mobility helps refuel the number of infecteds in a cluster. As $\lambda$ increases, the interplay of these two yields a critical value of $\lambda_c\approx 0.15$. A maximal severity now emerges at finite $\gamma_m$ obeying the approximate relationship $\gamma_m\sim e^3 q_i$. For a given $q_i$, the critical value of $\gamma_m$ separates a low-$\gamma_m$ phase in which increasing nodal mobility yields a decrease in outbreak severity, and a high-$\gamma_m$ phase in which increasing $\gamma_m$ yields an increase in severity. For $\lambda> 0.15$, the second mechanism dominates and there is a monotonic nonlinear increase in severity as $\gamma_m$ increases for all $q_i$.
The SM shows more details of the infection profiles, including the appearance of resurgent peaks with  quasi-oscillatory behavior (see SM). 
$A$, $T$ and $T_m$ tend to be maximal for smaller values of $q_i$, reflecting the slower spreading and hence longer duration. As $\lambda$ increases, the duration, time-to-peak and area become independent of mobility, because the well-mixed limit is approaching and the existence of clusters becomes unimportant (see SM). 

Figure 2(b) compares this variability and saturation effect in $T_m$  for one- and two-cluster versions. For the two-cluster version here, we choose the first cluster to have person-to-person infection while the second is broadcast, i.e. the individual infection rate is constant for every susceptible in this cluster at a given time: e.g. the first cluster mimics individuals in an online chatroom community while the second mimics individuals listening to the same radio broadcast. Two-cluster combinations can produce a larger ratio $T_m/T$ when compared with the one-cluster version, specifically $T_m/T>0.5$ as observed in the empirical data (Fig. 3(b)). 
Interestingly, 
the two-cluster series combination in Fig. 2(b) yields a near constant ratio $T_m/T$ for small values of mobility $\gamma_m$ but $T_m$ saturates as $T$ increases for larger $\gamma_m$, whereas the one-cluster version shows the opposite trend. 
Figure 2(c) shows that other disease processes (e.g. SIS) also display strong nonlinear dependences on nodal mobility $\gamma_m$ (Fig. 2(c)).

Figure 3(a) shows that the output from the nodal migration model compares favorably with empirical data for on-street (circles) and online (triangles) outbursts. While we are not suggesting it provides a unique or definitive explanation of these phenomena, the model (thin colored lines) does  capture the wide variability of outbreak profiles in a way that a standard SIR model cannot (thick black line). 
The on-street civil unrest data (circles) come from a unique multi-year, national research project involving exhaustive event analysis by subject matter experts (SMEs) across an entire continent (see Refs. \cite{unrest,IARPA}). The start and end of each burst is identified using the analysis of Ref. \cite{Karsai2} and cross-checked manually by SMEs. The online data comes from analysis of a European Facebook-like social media site which has attracted outbursts of pro-ISIS support through ad-hoc online communities whose followers rise and fall as a likely result of social contagion (see SM). 
The inset illustrates two of the underlying infection profiles ($I(t)$) with the corresponding color triangles. Figure 3(b) shows how the time-to-peak ($T_m$) and duration ($T$) of civil unrest outbreaks (color dots) relates to those generated from our model. The single cluster model captures outbreaks where $T_m<T/2$ (see Fig. 3(a), and middle and bottom simulation curve in Fig. 3(b)) while $M=2$ clusters in series extends the model's descriptive range to $T_m\rightarrow T$ in agreement with the data (Fig. 3(b) main panel).

Clusters in parallel (Fig. 1(c)) or series (Fig. 1(d)) mimic individuals who access one type of `space' such as a Facebook community, either at the same time as they check another (parallel case) or before they check another (series case). Figure 4 illustrates the rich infection profile behavior $I(t)$ that emerges for parallel (left column) and series (right column) clusters with person-to-person contagion. For the {\em parallel} case, we make the simple choice that an agent joins clusters $C_1$ or $C_2$ with probabilities $\omega_{p}$ and $1-\omega_{p}$ respectively. Hence in the steady state, $\left<N_{i}\right>=\gamma_{s,i}N$, for $i=1,2$, with $\gamma_{s,1}=\frac{p_{j}}{p_{j}+p_{l}}\omega_{p}$ and 
$\gamma_{s,2}=\frac{p_{j}}{p_{j}+p_{l}}\left(1-\omega_{p}\right)$. The $M=1$ case is recovered as $\omega_{p}\rightarrow 1$ or $\omega_{p}\rightarrow 0$. Figures 4(b) and (c) show that the infection peak height $H$ decreases significantly as $\omega_{p}\rightarrow 1/2$, while Fig. 4(a) shows a local maximum at $\omega_{p}=1/2$. These behaviors are favored by the average size of each cluster becoming similar as $\omega_{p}\rightarrow 1/2$. For the {\em series} case, we make the simple choice that an agent in $C_1$ joins cluster $C_2$ with probability $\omega_{s}$ and so on for $M>2$. An agent in the final cluster $C_M$ leaves it with probability $p_{l}$. Hence in the steady state for $M=2$, the mean number of nodes in $C_1$ and $C_2$ respectively is:
\begin{eqnarray}
\left<N_{1}\right>&=&N\gamma_{s}\left(\frac{\omega_{s}}{p_{l}}+\frac{p_{j}}{p_{l}+p_{j}}\right)^{-1}\nonumber\\
\left<N_{2}\right>&=&N\gamma_{s}\left({1+\frac{p_{l}}{\omega_{s}}\frac{p_{j}}{p_{j}+p_{l}}}\right)^{-1}\ .
\label{popS}
\end{eqnarray}
For $\omega_{s}\ll p_{l}$, $\left<N_{1}\right>\rightarrow N/(1+\kappa)$, where $\kappa=\omega_{s}/ p_{j}$, while $\left<N_{2}\right>\rightarrow 0$. By contrast for $\omega_{s}\gg p_{l}$, $\left<N_{1}\right>\rightarrow 0$ while for $C_{2}$ we recover the equilibrium population for the single cluster version (Fig. 1(a)). The asymmetry in Fig. 4 for $M=2$ clusters in series is strikingly different from the symmetry shown for the parallel case. This asymmetry has its roots in the breaking of symmetry in time (i.e. a node passes through $C_1$ before $C_2$). The infection profiles for the series case experience their largest variation for small $\omega_{s}$, with infection peaks that are significantly higher than for larger $\omega_{s}$. 
This is because at low $\omega_{s}$, 
$C_1$ has many nodes on average and these nodes are more likely to get infected and hence infect others. As $\omega_{s}$ grows and the size of $C_2$ approaches the $M=1$ case while the size of $C_1$ falls to zero, the infection profiles become identical to the $M=1$ case with $\omega_{s}\gg p_{l}$. There are significant differences in the infection profiles for broadcast transmission within a cluster as compared to person-to-person (see SM). This suggests distinct containment policies should be explored for outbreaks whose root cause is infected transient individuals (e.g. hospital patients or airline travelers) as opposed to infected transient places (e.g. the hospital or airport itself).

The general case of $M>2$ clusters allows for an interesting connection between the nodal migration dynamics and electric circuits (Fig. 1(e)) and a novel renormalization. Defining $k$ as the cluster label, $p_k$ as the probability to leave cluster $k$, and $n_k$ as the number of nodes in cluster $k$, we can associate an effective resistance $R_k\equiv 1/p_k$,  potential difference $U_k\equiv n_k$ and current $i_k\equiv \Delta n_k=U_k/R_k=n_k p_k$. This equivalence allows us to then generalize our model to $M$ clusters connected either in series or in parallel and hence quantify its dynamics. We have established this mapping exactly for $M>2$ clusters that are either all in series or in parallel (see SM). As an illustration in the steady state, the number of nodes in each of $M$ clusters connected either in series or in parallel is as follows: 
\begin{equation*}
\begin{aligned}
n_{0}^{(s)}&=\frac{N}{p_0}\left(\sum_{i=0}^{M}{\frac{1}{p_i}}\right)^{-1} & n_{0}^{(p)}&=\frac{N}{p_0}\left(\frac{1}{p_0}+\sum_{j=1}^{M}\frac{p_{0,j}}{p_0p_j}\right)^{-1}\\
n_{k}^{(s)}&=\frac{N}{p_k}\left(\sum_{i=0}^{M}{\frac{1}{p_i}}\right)^{-1} & n_{k}^{(p)}&=\frac{Np_{0,k}}{p_0 p_k}\left(\frac{1}{p_0}+\sum_{j=1}^{M}\frac{p_{0,j}}{p_0p_j}\right)^{-1}
\end{aligned}
\end{equation*}
where $k=0$ represents the nodes outside the fully connected set of clusters; $N$ is the total number of nodes; $p_{0,j}$ is the probability of moving from cluster $0$ to cluster $j$. Superscripts $s$ and $p$ denote series and parallel cases. For the series case, we can then regard the first $(M-1)$ clusters as a renormalized {\em super-cluster} $1'$ and replace the last cluster by cluster $2'$ with the following steady-state populations:
\begin{eqnarray}
\left<N'_{1}\right>&=&N\gamma_{s}\left(\frac{\omega'_{s}}{p_{l}}+\frac{p_{j}}{p_{l}+p_{j}}\right)^{-1}\nonumber\\
\left<N'_{2}\right>&=&N\gamma_{s}\left({1+\frac{p_{l}}{\omega'_{s}}\frac{p_{j}}{p_{j}+p_{l}}}\right)^{-1}.
\label{popS}
\end{eqnarray}
where $\omega'_{s}=(\sum_{j=1}^{M-1} \omega_{j}^{-1})^{-1}$ is the effective probability of nodes from cluster $1'$ migrating to cluster $2'$. $\omega_{j}$ is the migration probability from cluster $j$ to adjacent node $j+1$ in series. With this renormalization, effective two-cluster differential equations can then be written down and solved for the general $M$ case. 

In summary we have shown that nodal migration through a network generates highly complex outbreak profiles, even though the network appears static on average. We also indicated how the complex throughput of nodes can be renormalized exactly for a particular class of dynamical network. 

We are extremely grateful to Andrew Gabriel for help with the ISIS data collection, and to Chaoming Song and Stefan Wuchty for detailed discussions.

\end{document}